\documentclass[12pt]{article}
\usepackage{amssymb,amsmath,comment,mathtools}
\usepackage{graphicx}
\usepackage{xcolor}
\usepackage{hyperref}
\usepackage{cite}
\usepackage{subfigmat}
\usepackage{braket}
\usepackage{hyperref} 	  		
\usepackage{epsfig}
\usepackage{here}
\usepackage{xcolor}
\hypersetup{
    colorlinks=false,
    citebordercolor=green,
    linkbordercolor=red,
    urlbordercolor=cyan,
}
\usepackage{bm}
\graphicspath{{./Figures/}}

\newcommand{\ba}{\begin{align}}
\newcommand{\ea}{\end{align}}

\def\bea{\begin{eqnarray}}
\def\eea{\end{eqnarray}}
 \makeatletter
\def\alt{\mathrel{\mathpalette\gl@align<}}
\def\agt{\mathrel{\mathpalette\gl@align>}}
\def\gl@align#1#2{\lower.6ex\vbox{\baselineskip\z@skip\lineskip\z@
\ialign{$\m@th#1\hfil##\hfil$\crcr#2\crcr\sim\crcr}}} \makeatother
\renewcommand{\thefootnote}{\fnsymbol{footnote}}

\newcommand{\hc}{\mathrm{h.c.}}

\begin{document}
\begin{flushright}
{\small
NITEP 218}
\end{flushright}
\vspace*{1.0cm}

\begin{center}
\baselineskip 20pt 
{\Large\bf 
%Gauge Coupling Unification and 
Proton Decay and Gauge Coupling Unification in an Extended SU(5) GUT with 45-Dimensional Higgs}
\vspace{1cm}

{\large 
Naoyuki Haba${}^{a,b}$, Keisuke Nagano${}^{a,c}$, Yasuhiro Shimizu${}^{a,b}$ \\
and Toshifumi Yamada${}^{d}$
} \vspace{.5cm}

{\baselineskip 20pt \it
${}^{a}$Department of Physics, Osaka Metropolitan University, Osaka 558-8585, Japan \\
${}^{b}$Nambu Yoichiro Institute of Theoretical and Experimental Physics (NITEP),
Osaka Metropolitan University, Osaka 558-8585, Japan\\
${}^{c}$Institute of Science and Engineering, Shimane University, Matsue 690-8504, Japan\\
${}^{d}$Institute for Mathematical Informatics, Meiji Gakuin University, Yokohama 244-8539, Japan
}

\vspace{.5cm}

\vspace{1.5cm} {\bf Abstract} \end{center}

We present a comprehensive study of an extended SU(5) grand unified theory (GUT) that incorporates a 45-dimensional Higgs representation to address the shortcomings of the minimal SU(5) GUT, such as the inability to generate realistic fermion mass hierarchies and insufficient proton stability. By considering a hierarchical mass spectrum for the scalar components of the 45-Higgs, we demonstrate that successful gauge coupling unification (GCU) can be achieved. The color octet scalar, color triplet scalar, and color anti-triplet scalar play crucial roles in realizing GCU when their masses are significantly lighter than other components of the 45-Higgs. 
We focuses on the proton decay channels mediated by the exchange of the color anti-triplet scalar. Assuming that the 45-Higgs couples to all three generations of fermions, we determine the 45-Higgs Yukawa couplings with which the observed fermion mass matrices at low energies are realized. We calculate proton decay rates using the Yukawa couplings obtained from renormalization group evolutions and matching conditions at the GUT scale, thereby exploring the dependence of proton decay rates on model parameters. We find that the 
$p \to  \nu \pi$ mode imposes the most stringent constraint on the mass of the color anti-triplet scalar $M_{S_1}$.
We also study the correlations between the lower bounds on $M_{S_1}$ derived from different proton decay modes.

\thispagestyle{empty}

%\bigskip
\newpage
\renewcommand{\thefootnote}{\arabic{footnote}}
%\addtocounter{page}{-1}
\setcounter{footnote}{0}
%%%%%%%%%%%%%%%%%%%%%%%%%%
%\baselineskip 36pt
% Main body
%%%%%%%%%%%%%%%%%%%%%%%%%%
\baselineskip 18pt
%%%%%%%%%%%%%%%%%%%%%%%%%%

\section{Introduction}

The Standard Model (SM) of particle physics has been remarkably successful in describing phenomena at currently accessible energy scales. However, it leaves several fundamental unanswered questions, such as the origin of fermion masses and mixings, the quantization of hypercharge, and the unification of the three gauge interactions. Grand Unified Theories (GUTs) provide an appealing framework to address these issues by embedding the SM gauge group into a larger simple group~\cite{GUT1}-\cite{GUT6}.

One of the most studied GUT models is based on the SU(5) gauge group, which allows the unified description of the strong and electroweak interactions~\cite{GUT3}. Despite its elegance, the minimal SU(5) model faces several difficulties. It predicts the same Yukawa couplings for down-type quarks and charged leptons at the GUT scale, which contradict  the observed mass relations. Furthermore, the three gauge couplings do not precisely unify at a single high-energy scale in non-supersymmetric models.

To overcome these limitations, extensions to the minimal SU(5) GUT have been proposed. One of the promising approaches is inclusion of a special type of Higgs field. If the Higgs field transforming as the 45-dimensional representation is introduced, the 45-dimensional Higgs can generate separate Yukawa matrices for quarks and leptons, potentially accommodating the observed fermion mass hierarchies and mixing patterns~\cite{45-1}-\cite{GeorgiJarlskog}. Furthermore, the GCU can achieved if we consider hierarchical mass spectrum for the components of the 45 Higgs.
In Ref.~\cite{Haba:2024lox},  we  studied the GCU and proton decay via 45-dimensinal Higgs boson focused on the 45 Higgs coupling to the second generation, to explain specific mass ratios like that of the strange quark and muon through Georgi-Jarlskog-type mass matrices~\cite{GeorgiJarlskog}.

In this paper, we extend the analysis by considering an SU(5) GUT model where the 45-representation Higgs field couples to all three generations of fermions. The Yukawa couplings of the 45 Higgs are determined such that the full structure of the observed fermion mass matrices, including hierarchies and mixing angles, can be reproduced at low energies. 
To achieve successful GCU and satisfy current experimental constraints on proton decay, we introduce a hierarchical mass spectrum for the components of the ${45}_H$ Higgs multiplet. By solving the renormalization group equations (RGEs) with various mass hierarchies among the ${45}_H$ components, we identify viable parameter regions that lead to precise unification of the gauge couplings while evading stringent experimental bounds on proton decay.
We focus on the case where the three components of the 45-Higgs, which are the ($\overline{3}$,1), (3,3), and (8,2) representations of the SM gauge group, are much lighter than the GUT scale. We study proton decay mediated by the color anti-triplet Higgs components and derived lower mass bounds on these components from the latest experimental constraints, such as those set by the Super-Kamiokande experiment. By analyzing the interplay among the Yukawa couplings, the mass hierarchies of the ${45}_H$ Higgs components, and the RGE evolution of the gauge couplings,  we find that the 
$p \to  \nu \pi$ mode imposes the most stringent constraint on the color anti-triplet Higgs mass. We also study the correlations among the lower bounds on the color anti-triplet Higgs mass  derived from different proton decay modes.

This paper is organized as follows. In Sect. 2, we review the SU(5) GUT with a 45 Higgs field. In Sect. 3, we discuss the GCU conditions assuming the components of the 45 Higgs field have a hierarchical mass spectrum. In Sect. 4, we study the contribution of the colored Higgs component of the 45 Higgs field to proton decay and identify the parameter region allowed by all the experiments. Section 5 concludes the paper.

%%%%%%%%%%%%%%%%%%%%%%%%%%%%%%%%%%%%%%%%%%
%%%               Model              %%%%%
%%%%%%%%%%%%%%%%%%%%%%%%%%%%%%%%%%%%%%%%%%
\section{SU(5) GUT with {\bf 45} dimensional Higgs}

In this section, we consider a non-SUSY SU(5) GUT with a $\bm{45}$ Higgs boson. The fermions in the SM are embedded into the $\bm{\bar{5}}$ and $\bm{10}$ representations of SU(5) as follows:
\begin{align}
\bm{\bar{5}}: \overline{\psi}_{\hat{a}i}
=
\begin{pmatrix}
d_{Rai}^{C} &
\epsilon_{\alpha\beta}\, (V_{DL})_i{}^j\,
\ell_{Lj}^{\, \beta} 
\end{pmatrix},~~~~
\bm{10}:\chi_{i}^{\hat{a}\hat{b}} =
\frac{1}{\sqrt{2}}
\begin{pmatrix}
\epsilon^{abc}\, (V_{QU})_i{}^j\,
u^{C}_{Rcj} &
q_{Li}^{\, a\beta}
\\[1mm]
-q_{Li}^{\, b\alpha} &
\epsilon^{\alpha\beta}\, (V_{QE})_i{}^j\,
e_{Rj}^{\, C}
\end{pmatrix}.
\end{align}
Here, $\hat{a},\hat{b}$ are SU(5) indices, $a,b,c$ are SU(3) indices, $\alpha,\beta$ are SU(2) indices, and $i,j$ are generation indices. The unitary matrices $V_{DL}$, $V_{QU}$, and $V_{QE}$ are determined by imposing matching conditions between the GUT and SM fermion sectors at the unification scale.

While the minimal SU(5) GUT incorporates Higgs transforming as the $\bm{24}$ and $\bm{5}$ representations, we extend this by introducing an additional Higgs multiplet transforming as the $\bm{45}$ representation of SU(5), denoted as ${45}_H$. The inclusion of ${45}_H$ allows for the generation of realistic fermion mass patterns and mixings, which cannot be fully accounted for by the minimal Higgs representations. The SU(5) invariant Yukawa interactions can be written as:
\begin{align}
-\mathcal{L}_Y
  &=
  \frac{1}{8}\, (Y_5^U)_{ij}\epsilon_{\hat{a}\hat{b}\hat{c}\hat{d}\hat{e}}\,
    \chi_{i}^{\hat{a}\hat{b}}\,\chi_{j}^{\hat{c}\hat{d}}\,(5_H)^{\hat{e}}
  + (Y_5^D)_{ij}
    \chi_{i}^{\hat{a}\hat{b}}\, \overline{\psi}_{\hat{a}j} \,(5_H^\dagger)_{\hat{b}}
  \nonumber\\
  &+ \frac{1}{4}(Y_{45}^U)_{ij}\epsilon_{\hat{a}\hat{b}\hat{c}\hat{d}\hat{e}}\,
    \chi_{i}^{\hat{a}\hat{b}}\,\chi_{i}^{\hat{c}\hat{f}}\,({45}_H)^{\hat{d}\hat{e}}_{\hat{f}}
  + \frac{1}{2}\, (Y_{45}^D)_{ij}
    \chi_{i}^{\hat{a}\hat{b}}\,\overline{\psi}_{\hat{c}j}({45}_H^\dagger)_{\hat{a}\hat{b}}^{\hat{c}}
  + \mathrm{h.c.}.
    \label{GUTYukawa}
\end{align}
Notice that ${45}_H$  are assumed to couple with all three generations of fermions. By choosing these Yukawa coupling matrices, we can reproduce the observed patterns of fermion masses and mixings. The 45 Higgs field, in particular, plays a crucial role in generating realistic mass relations and mixing angles that cannot be accounted for by the minimal Higgs representations alone.

The ${5}_H$ and ${45}_H$ Higgs fields can be decomposed into their components according to the SM representations as follows:
\begin{align}
{5}_H &=H^{(5)} (1,2)_{\frac{1}{2}}
 \oplus S_1^{(5)*} (3,1)_{-\frac{1}{3}},
 \\
{45}_H &=H^{(45)} (1,2)_{\frac{1}{2}}
    \oplus S_1^{(45)*} (3,1)_{-\frac{1}{3}}
    \oplus S_3^* (3,3)_{-\frac{1}{3}}
    \nonumber\\
    &\oplus  \tilde{S}_1(\overline{3},1)_{\frac{4}{3}}    
    \oplus R_2^* (\overline{3},2)_{-\frac{7}{6}}
    \oplus S_6^* (\overline{6},1)_{-\frac{1}{3}}
    \oplus S_8 (8,2)_{\frac{1}{2}}.
\end{align}
In order to achieve the GCU, we consider the mass hierarchy among the components in the ${45}_H$ Higgs and assume that some components are much lighter than the GUT scale.

Below the GUT scale, the $H^{(45)}(1,2)_{\frac{1}{2}}$ and $S_1^{(45)}(3,1)_{-\frac{1}{3}}$ components mix with the $H^{(5)}(1,2)_{\frac{1}{2}}$ and $S_1^{(5)}(3,1)_{-\frac{1}{3}}$, respectively, 
depending on the parameters of the scalar potential.
The mass eigenstates resulting from these mixings are given as follows:
\begin{align}
  \begin{pmatrix}
  H\\ H^{\prime}
  \end{pmatrix}
  = 
  \begin{pmatrix}
  c_H & e^{-i\delta_H}s_H\\
  -e^{i\delta_H}s_H & c_H 
  \end{pmatrix}
  \begin{pmatrix}
  H^{(5)}\\ H^{(45)}
  \end{pmatrix},
  \qquad
  \begin{pmatrix}
  H_C\\ S_1
  \end{pmatrix}
  = 
  \begin{pmatrix}
  c_S & e^{-i\delta_S} s_S\\
  -e^{i\delta_S}s_S & c_S 
  \end{pmatrix}
  \begin{pmatrix}
  S_1^{(5)}\\ S_1^{(45)}
  \end{pmatrix},
\label{eq:mixing}
\end{align}
where 
$c_H=\cos\theta_H$, $s_H=\sin\theta_H$,
$c_S=\cos\theta_S$, and $s_S=\sin\theta_S$.\footnote{We have assumed that  the colored Higgs in the 5-dimensional Higgs boson, $S_1^{(5)}(3,1)_{-\frac{1}{3}}$, has a mass around the GUT scale and the lighter mass eigenstate, $S_1$, is primarily composed of $S_1^{(45)}(3,1)_{-\frac{1}{3}}$ with a tiny mixing angle $\theta_S\sim m^2_{S_1^{(45)}}/m^2_{S_1^{(5)}}\ll 1$. }
The SU(5) invariant Yukawa interactions in Eq.(\ref{GUTYukawa})
can be  decomposed in terms of the component
fields as follows~\cite{Goto:2023qch}:
\begin{align}
- \mathcal{L}_Y 
 & =
  (Y_U)_{ij}\epsilon_{\alpha\beta}\,
    \bar{u}_{R a i}\, H^{\alpha}q^{a\beta}_{Lj}
  + (Y_D)_{ij}\,
    \bar{d}_{Ra i}\, H_{\alpha}^*\, q^{a\alpha}_{Lj}
  + (Y_E)_{ij}\,
    \bar{e}_{Ri}\, H_{\alpha}^*\, \ell^{\,\alpha}_{Lj}
  \nonumber\\
  & 
  + (Y_U^{\prime})_{ij}\epsilon_{\alpha\beta}\,
    \bar{u}_{R a i}\, H^{\prime\alpha} q^{a\beta}_{Lj}
  + (Y_D^{\prime})_{ij}\,
    \bar{d}_{R a i}\, H_{\alpha}^{\prime *}\, q^{a\alpha}_{Lj}
  + (Y_E^{\prime})_{ij}\,
    \bar{e}_{Ri}\, H_{\alpha}^{\prime *}\, \ell^{\,\alpha}_{Lj}
  \nonumber\\
  & 
  + (Y_C^{QL})_{ij}\epsilon_{\alpha\beta}\,
    \bar{q}^{\, C\, a\alpha}_{Li} H_{C\, a}\, \ell_{Lj}^{\,\beta}
  + (Y_C^{UE})_{ij}\,
    \bar{u}^{}_{R\, a i}\, H_C^{*a} e_{Rj}^C
  + (Y_C^{DU})_{ij}\epsilon^{abc}
    \bar{d}_{R\,a i}\, H_{C\,b}\, u_{R\, c j}^{C}
      \nonumber\\
  &
  + \frac{(Y_C^{QQ})_{ij}}{2}\,\epsilon_{abc}\,\epsilon_{\alpha\beta}\,
    \bar{q}^{C\,a\,\alpha}_{Li}H_C^{*\, b} q_{Lj}^C\,\beta
      + \frac{(Y_1^{QQ})_{ij}}{2}\,\epsilon_{a b c}\,\epsilon_{\alpha\beta}\,
    \bar{q}^{\, C\,a\alpha}_{Li}S_1^{*\hat{b}} q_{Lj}^{C\beta}
  \nonumber\\
  &
  + (Y_1^{QL})_{ij}\epsilon_{\alpha\beta}\,
    \bar{q}^{\, C\,a\alpha}_{Li} S_{1a}\, \ell_{Lj}^{\,\beta}
  + (Y_1^{UE})_{ij}\,
    \bar{u}^{}_{Ra i}\, S_1^{*a} e_{Rj}^C
  + (Y_1^{DU})_{ij}\epsilon^{a b c}\,
    \bar{d}_{Ra i}\, S_{1b}\, u_{Rc j}^C
  \nonumber\\
  &
  + (\tilde{Y}_1^{ED})_{ij}\,
    \bar{e}_{Ri}^{}\, \tilde{S}_{1}^{*a}d_{Ra j}^{\, C}
  + \frac{(\tilde{Y}_1^{UU})_{ij}}{2}\,\epsilon^{a b c}\,
    \bar{u}_{Ra i}\,\tilde{S}_{1b}\, u_{Rc j}^C
  + (Y_2^{UL})_{ij}\epsilon_{\alpha\beta}\,
    \bar{u}_{Ra i}\, R_{2}^{\,a\alpha}\ell_{Lj}^{\,\beta}
  + (Y_2^{EQ})_{ij}\,
    \bar{e}_{Ri}\, R_{2a\alpha}^{\, *}\, q_{Lj}^{a\alpha}
  \nonumber\\
  &
  + (Y_3^{QL})_{ij}\epsilon_{\alpha\beta}\,
    \bar{q}_{Li}^{\, C\,a\gamma}(\sigma_A)^{\alpha}{}_{\gamma}\,
    S_{3a}^A\,\ell_{Lj}^{\,\beta}
  + \frac{(Y_3^{QQ})_{ij}}{2}\,
    \epsilon_{a b c}\,\epsilon_{\alpha\beta}\,
    \bar{q}_{Li}^{C\,a\alpha}(\sigma_A)^{\beta}{}_{\gamma}\,
    S_3^{* Ab}q_{Lj}^{C\,\gamma}
  \nonumber\\
  &
  + (Y_6^{DU})_{ij}\,
    \bar{d}_{Ra i}\, (\eta_A)^{a b}
    S_{6}^{A}\, u_{Rb j}^C
  +\frac{(Y_6^{QQ})_{ij}}{2}\,\epsilon_{\alpha\beta}\,
    \bar{q}_{Li}^{\, C\,a\alpha} (\eta_A)_{a b}\, 
    S_{6}^{A*}\, q_{Lj}^{b\beta}
  \nonumber\\
  &
  + (Y_8^{UQ})_{ij}\epsilon_{\alpha\beta}\,
    \bar{u}_{Ra i}^{}\, (\lambda_A)^{a}{}_{b}\,
    S_{8}^{A \alpha }\, q_{Lj}^{b\beta}
  + (Y_8^{DQ})_{ij}\,
    \bar{d}_{Ra i}\, (\lambda_A)^{a}{}_{b}\,
    S_{8\alpha}^{A*}\, q_{Lj}^{b\alpha}
  + \hc\,,    
\end{align}
Here the above Yukawa couplings are matched to the GUT Yukawa couplings at the GUT scale as follows:
\begin{align}
  Y_U =&
    -\frac{1}{2}\, V_{QU}^{T}
    \left( c_H\, Y_{5}^U 
      + \sqrt{\frac{2}{3}}\,e^{i\delta_H}s_H\, Y_{45}^{U}\right)^{ T},
  &
  Y_U^{\prime} =& 
    \frac{1}{2}\, V_{QU}^{ T}
    \left(e^{-i\delta_H}s_H\, Y_{5}^U
    - \sqrt{\frac{2}{3}}\,c_H\, Y_{45}^{U}\right)^{ T},
  \nonumber\\ 
  Y_D =& 
    - \frac{1}{\sqrt{2}}
    \left( c_H\, Y_{5}^{D}
    - \frac{1}{2\sqrt{6}}\,e^{-i\delta_H}s_H\, Y_{45}^{D} \right)^{ T},
  &
  Y_D^{\prime} =&
    \frac{1}{\sqrt{2}}
    \left( e^{i\delta_H}s_H\, Y_{5}^{D}
    + \frac{1}{2\sqrt{6}}\,c_H\, Y_{45}^{D} \right)^{ T},
  \nonumber\\ 
  Y_E =&
    - \frac{1}{\sqrt{2}}\, V_{QE}^T
    \left( c_H\, Y_{5}^D
    + \frac{\sqrt{3}}{2\sqrt{2}}\,e^{-i\delta_H}s_H\, Y_{45}^{D}\right)
    V_{DL}\,,
  &
  Y_E^{\prime} =&
    \frac{1}{\sqrt{2}}\, V_{QE}^T
    \left(e^{i\delta_H}s_H\, Y_{5}^{D}
    - \frac{\sqrt{3}}{2\sqrt{2}}\, c_H\, Y_{45}^{D}\right)
    V_{DL}\,,
  \nonumber\\
  Y_C^{QL} =&
    \frac{1}{\sqrt{2}}
    \left(
    c_S\, Y_{5}^{D}
    + \frac{1}{2\sqrt{2}}\,e^{i\delta_S}s_S\, Y_{45}^{D}\right)
    V_{DL}\,,
  &
  Y_1^{QL} =&
    \frac{1}{\sqrt{2}}
    \left(
    - e^{-i\delta_S}s_S\, Y_{5}^{D}
    + \frac{1}{2\sqrt{2}}\,c_S\, Y_{45}^{D} \right) 
    V_{DL}\,,
  \nonumber\\
  Y_C^{UE} =&
    \frac{1}{2}\, V_{QU}^{\, T}
    \left( c_S\, Y_{5}^{U}
    - \sqrt{2}\,e^{-i\delta_S}s_S\, Y_{45}^{U} \right)
    V_{QE}\,,
  &
  Y_1^{UE} =&
    - \frac{1}{2}\, V_{QU}^{ T}
    \left( e^{i\delta_S}s_S\, Y_{5}^{U}
    + \sqrt{2}\, c_S\, Y_{45}^{U} \right)
    V_{QE}\,,
  \nonumber\\
  Y_C^{DU} =&\ 
    \frac{1}{\sqrt{2}} \left(
    - c_S\, Y_{5}^{D}
    + \frac{1}{2\sqrt{2}}\, e^{i\delta_S}s_S\, Y_{45}^{D}
    \right)^{ T}
    V_{QU}\,,
  &
  Y_1^{DU} =&
    \frac{1}{\sqrt{2}} \left(
    e^{-i\delta_S}s_S\, Y_{5}^{D}
    + \frac{1}{2\sqrt{2}}\, c_S\, Y_{45}^{D}
    \right)^{ T}
    V_{QU}\,,
  \nonumber\\
  Y_C^{QQ} =&\ 
    \frac{1}{2}\, c_S\, Y_{5}^{U},
~~
  Y_1^{QQ} =
    - \frac{1}{2}\, e^{i\delta_S}s_S\, Y_{5}^{U},
  &
  \tilde{Y}_1^{UU} =&
    \frac{1}{\sqrt{2}}\, V_{QU}^{\, T}\, Y_{45}^{U}\,
    V_{QU}\,,
~~
  \tilde{Y}_1^{ED} =
    \frac{1}{2}\, V_{QE}^{ T}\, Y_{45}^{D}\,,
  \nonumber\\
  Y_2^{EQ} =&
    \frac{1}{\sqrt{2}}\, V_{QE}^{ T}\,Y_{45}^{U}\,,
  \qquad
  Y_2^{UL} =
    \frac{1}{2}\, V_{QU}^{T}\, Y_{45}^{D}\, 
    V_{DL}\,,
  &
  Y_3^{QQ} =&
    \frac{1}{2}\, Y_{45}^{U}\,,
~~~
  Y_3^{QL} =
    - \frac{1}{2\sqrt{2}}\, Y_{45}^{D}\, V_{DL}\,,
  \nonumber\\
  Y_6^{QQ} =&
    - \frac{1}{\sqrt{2}}\, Y_{45}^{U}\,,
~~
  Y_6^{DU} =
    \frac{1}{2}\, (Y_{45}^{D})^{T}\,  V_{QU}\,,
  &
  Y_8^{UQ} =&
    - \frac{1}{2}\, 
    V_{QU}^{ T}\,
    Y_{45}^{U}\,,
~~
  Y_8^{DQ} = 
    \frac{1}{2\sqrt{2}}\, (Y_{45}^{D})^{T}.
    \label{eq:Ymatching}
\end{align}
If we take the basis where the up-type quarks and the charged
leptons are their mass eigenstates, the GUT Yukawa matrices $Y_{5}^{U}$, $Y_{45}^{U}$,
$Y_5^{D}$, and $Y_{45}^{D}$ can be written as follows: 
\begin{align}
Y_{5}^U
  &=
  -\frac{1}{c_H}
  \left(\, V_{QU}^*\, \hat{Y}_U 
  + \hat{Y}_U\, V_{QU}^\dagger \right)\,,
&Y_{45}^U
  &=
  \frac{\sqrt{3}}{\sqrt{2}\,e^{i\delta_H}s_H}
  \left(\, V_{QU}^*\, \hat{Y}_U 
  - \hat{Y}_U\, V_{QU}^\dagger \right)\,,
  \nonumber\\ 
Y_{5}^{D}
  &= 
  - \frac{1}{2\sqrt{2}\, c_H}
  \left( 3\, V_{\text{CKM}}^*\, \hat{Y}_D
  + V_{QE}^*\, \hat{Y}_E\, V_{DL}^{\dagger}
  \right)\,,
&Y_{45}^{D}
  &=
  \frac{\sqrt{3}}{e^{-i\delta_H}s_H}
  \left(
  V_{\text{CKM}}^{*}\, \hat{Y}_D
  - V_{QE}^*\, \hat{Y}_E\, V_{DL}^{\dagger}
  \right)\,,
     \label{eq:Ymatching2}
\end{align}
where $\hat{Y}_U$, $\hat{Y}_D$, and $\hat{Y}_E$ represent diagonal
matrices in the mass basis and $V_{\text{CKM}}$ is the Cabibbo-Kobayashi-Maskawa matrix, which
are determined by the observed fermion masses and mixing angles. On the other hand, the unitary matrices, $V_{QE}$, $V_{QU}$, $V_{DL}$, cannot be determined solely from low-energy experiments. These matrices represent the rotations required to diagonalize the quark-Higgs and lepton-Higgs Yukawa couplings in the SU(5) GUT framework and are important for the proton decay analysis.

\section{Gauge Coupling Unification}
In this section we briefly review the result of GCU in SU(5) GUT with 45-Higgs discussed in Ref.~\cite{Haba:2024lox}. By solving the renormalization group equations (RGEs) at 1-loop, the running gauge coupling constants at an energy scale $Q$ can be expressed as follows:
\begin{align}
    \alpha_i^{-1}(M_Z)=\alpha_i^{-1}(Q)-\frac{\beta_i}{2\pi}\log\left( \frac{M_Z}{Q}\right).
\end{align}
Here $M_Z$ is the $Z$ boson mass and $\beta_i$'s represent the beta functions of the three gauge coupling constants. Assuming the hierarchical mass spectrum for the $45$ Higgs, the solutions 
are given as follows:
\begin{align}
	\alpha_1^{-1}(M_Z)=\alpha_1^{-1}(\Lambda)+\frac{1}{2\pi}\biggl[&\frac{41}{10}\log\frac{M_Z}{\Lambda}+\frac{1}{15}\log\frac{M_{H_C}}{\Lambda}-\frac{35}{2}\log\frac{M_{XY}}{\Lambda}+\frac{1}{10}\log\frac{M_{H^{(45)}}}{\Lambda}
 \nonumber\\
	&+\frac{1}{15}\log\frac{M_{S_1^{(45)}}}{\Lambda}+\frac{1}{5}\log\frac{M_{S_3}}{\Lambda}
	 +\frac{16}{15}\log\frac{M_{\tilde{S}_1}}{\Lambda}\nonumber\\
	&+\frac{49}{30}\log\frac{M_{R_2}}{\Lambda}
	 +\frac{2}{15}\log\frac{M_{S_6}}{\Lambda}
	 +\frac{4}{5}\log\frac{M_{S_8}}{\Lambda}\biggr]-\frac{5}{12\pi},
     \label{alpha1}\\
	\alpha_2^{-1}(M_Z)=\alpha_2^{-1}(\Lambda)+\frac{1}{2\pi}\biggl[&-\frac{19}{6}\log\frac{M_Z}{\Lambda}-\frac{21}{2}\log\frac{M_{XY}}{\Lambda}+\frac{1}{3}\log\frac{M_\Sigma}{\Lambda}+\frac{1}{6}\log\frac{M_{S_1}}{\Lambda}\nonumber\\
	&+2\log\frac{M_{S_3}}{\Lambda}
	 +\frac{1}{2}\log\frac{M_{R_2}}{\Lambda}
	 +\frac{4}{3}\log\frac{M_{S_8}}{\Lambda}\biggr]-\frac{5}{12\pi},
  \label{alpha2} \\
  	\alpha_3^{-1}(M_Z)=\alpha_3^{-1}(\Lambda)+\frac{1}{2\pi}\biggl[&-7\log\frac{M_Z}{\Lambda}-\frac{1}{6}\log\frac{M_{H_C}}{\Lambda}-7\log\frac{M_{XY}}{\Lambda}-\frac{1}{2}\log\frac{M_\Sigma}{\Lambda}\nonumber\\
	&+\frac{1}{6}\log\frac{M_{S_1^{(45)}}}{\Lambda}
	 +\frac{1}{2}\log\frac{M_{S_3}}{\Lambda}
	 +\frac{1}{6}\log\frac{M_{\tilde{S}_1}}{\Lambda}\nonumber\\
	&+\frac{1}{3}\log\frac{M_{R_2}}{\Lambda}
	 +\frac{5}{6}\log\frac{M_{S_6}}{\Lambda}
	 +2\log\frac{M_{S_8}}{\Lambda}\biggr]-\frac{5}{12\pi},
   \label{alpha3}
\end{align}
where $M_{XY}$, $M_\Sigma$ are the the messes of the $X$ gauge boson, 24-Higgs boson, respectively.
If the GCU are achieved at $\Lambda$, the following conditions are satisfied. 
\begin{align}
    \alpha_1(\Lambda)=\alpha_2(\Lambda)=\alpha_3(\Lambda),
\end{align}
From the Eqs.(\ref{alpha1})-(\ref{alpha3}) we can obtain the following equations.
\begin{align}
	\alpha_1^{-1}(M_Z)-3\alpha_2^{-1}(M_Z)+2\alpha_3^{-1}(M_Z)=-\frac{1}{2\pi}\biggl[&\frac{2}{5}\log\frac{M_{H_C}}{M_Z}-\frac{2}{5}\log\frac{M_{H^{(45)}}}{\Lambda}+\frac{2}{5}\log\frac{M_{S_1}}{\Lambda}\nonumber\\
	&-\frac{24}{5}\log\frac{M_{S_3}}{\Lambda}
	 +\frac{7}{5}\log\frac{M_{\tilde{S}_1}}{\Lambda}
	 +\frac{4}{5}\log\frac{M_{R_2}}{\Lambda}\nonumber\\
	&+\frac{9}{5}\log\frac{M_{S_6}}{\Lambda}
	 +\frac{4}{5}\log\frac{M_{S_8}}{\Lambda}\biggr]
	 \\
	-5\alpha_1^{-1}(M_Z)+3\alpha_2^{-1}(M_Z)+2\alpha_3^{-1}(M_Z)=-\frac{1}{2\pi}\biggl[&42\log\frac{M_{XY}}{M_Z}
	 +2\log\frac{M_\Sigma}{M_Z}
	 +6\log\frac{M_{S_3}}{\Lambda}\nonumber\\
	&-5\log\frac{M_{\tilde{S}_1}}{\Lambda}
	 -6\log\frac{M_{R_2}}{\Lambda}
	 +\log\frac{M_{S_6}}{\Lambda}\nonumber\\
	&+4\log\frac{M_{S_8}}{\Lambda}\biggr].
\end{align}

In Ref.~\cite{Haba:2024lox}, it has been shown that at least two light components of 45-Higggs are required in order to satisfy the GCU conditions. For the case of two light components, we found that proton decay via the two light components does not occur. For the three light components case where the components (3,1), (3,3), and (8,2) are light, the (3,1) component, i.e. $S_1$, induces proton decay and so we can constrain the model from the proton decay bounds.
To achieve GCU we consider the mass hierarchy where $S_8$, $S_3$, $S_1$ are lighter than the other components of $45$-Higgs. For example,
it has been demonstrated that GCU can be realized  
for the following scalar components masses:
\begin{align}
M_{S_8} = 3 \times 10^{4}~\mathrm{GeV}, \quad M_{S_3} = 5 \times 10^{7}~\mathrm{GeV}, \quad M_{S_1} = 1.8 \times 10^{12}~\mathrm{GeV}.
\label{eq:spec}
\end{align}
With this mass hierarchy, the GCU is achieved, and in this case the masses of the GUT gauge bosons and the other scalar particles are $6.5 \times 10^{15}~\mathrm{GeV}$.
Furthermore, it has been shown that GCU can be achieved in a wider region of the hierarchical mass spectrum where $S_8$, $S_3$, and $S_1$ are lighter than the other components of $45$-Higgs. 
%In particular, GCU is realized in the region satisfying:
%\begin{align}
% 2 \times 10^{13}\mathrm{GeV}\lesssim M_{S_3} \lesssim 2 \times 10^{18}\mathrm{GeV}.
%\label{eq:spec}
%\end{align}
%This range of $M_{S_3}$ values allows for a broader parameter space in which successful GCU can be achieved while maintaining the hierarchical mass spectrum of the scalar components.

In the next section, we focus on the proton decay mediated by the exchange of $S_1$. The proton decay widths and their dependence on the model parameters, such as the Yukawa couplings and the mass of $S_1$, will be investigated in detail.

\section{Proton decay}
When the  $S_1$, $S_3$, and $S_8$ components of the 45-Higgs are significantly lighter than the GUT scale, the proton decay can be induced via the  $S_1$ and $S_3$  exchange diagrams. As discussed in Ref.~\cite{Haba:2024lox}, the proton decay mediated by the $S_3$ exchange imposes stringent constraints due to the presence of two Yukawa couplings, $Y^{QQ}_3$ and $Y^{QL}_3$, which govern the interaction between the $S_3$ component and the quarks and leptons. These Yukawa couplings directly contribute to the proton decay amplitudes.
The current experimental limits from proton decay searches, specifically the bound on the process $p\to e^+\pi$, constrain the product of these Yukawa couplings as follows:
\begin{align}
|(Y_3^{QQ})_{12} (Y_3^{QL})_{11}| < 10^{-17} s_H,
\end{align}
for $m_{S_3}=5\times 10^7$ GeV. 
By invoking the matching conditions between the GUT and SM Yukawa couplings, as given in Eqs.~(\ref{eq:Ymatching}), the above bound can be translated into a constraint on the GUT Yukawa coupling associated with the 45-Higgs:
\begin{align}
|(Y^U_{45})_{12}| < 10^{-12} s_H.
\end{align}
To suppress the proton decay rate mediated by the $S_3$ component and satisfy the experimental constraints, we adopt the assumption that the GUT Yukawa coupling $Y^U_{45}$ is forbidden at the GUT scale, following the approach in Ref.~\cite{Goto:2023qch}. 
Even if  $Y^U_{45}$ is forbidden at the GUT scale, $Y_3^{QQ}$ can be generated
through a one-loop RGE effect involving $S_1$.  However, the corresponding term of the one-loop beta function of $Y_3^{QQ}$ is proportional to the product $(Y_1^{QL})(Y_3^{QL})^*(Y_1^{QQ})$. Since we have assumed that $S_1$ is predominantly composed of the 45-dimensional Higgs boson, $Y_1^{QQ}$ is highly suppressed by the tiny mixing angle $\theta_S$. 
Consequently, the one-loop contribution to $Y_3^{QQ}$ 
  is negligible, ensuring that the proton decay rate remains sufficiently suppressed.
Even with this assumption, it is possible to reproduce realistic fermion masses and mixings within the framework of the extended GUT model.
We focus our analysis on the proton decay process mediated by the exchange of the $S_1$ component. The partial decay widths for these processes are calculated using the effective Lagrangian and taking into account the relevant hadronic matrix elements and form factors as follows \cite{Nath:2006ut}:
\begin{align}
\Gamma(p\to\mu^+K^0)&=\frac{1}{64\pi}\left(1-\frac{m_K^2}{m_p^2}\right)^2\frac{m_p}{f^2} 
\left\{
 (1+D-F)^2\alpha_H^2A_{RL}^2 \frac{|\delta^L_{2112}|^2 }{M_{S_1}^4}
 + (1-D+F)^2\beta_H^2A_{RL}^2 \frac{|\delta^R_{2112}|^2 }{M_{S_1}^4}
 \right\},
\label{ptomuk}\\
\Gamma(p\to\mu^+\pi^0)&=\frac{1}{64\pi}\left(1-\frac{m_\pi^2}{m_p^2}\right)^2\frac{m_p}{f^2} \ \frac{1}{2}(1+D+F)^2
\left\{
 \alpha_H^2A_{RL}^2 \frac{|\delta^L_{1112}|^2 }{M_{S_1}^4}
+\beta_H^2A_{RL}^2 \frac{|\delta^R_{1112}|^2 }{M_{S_1}^4}
 \right\},
\label{ptomupi}\\
\Gamma(p\to e^+K^0)&=\frac{1}{64\pi}\left(1-\frac{m_K^2}{m_p^2}\right)^2\frac{m_p}{f^2} 
\left\{
 (1+D-F)^2\alpha_H^2A_{RL}^2 \frac{|\delta^L_{2111}|^2 }{M_{S_1}^4}
 + (1-D+F)^2\beta_H^2A_{RL}^2 \frac{|\delta^R_{2111}|^2 }{M_{S_1}^4}
 \right\},
\label{ptoek}\\
\Gamma(p\to e^+\pi^0)&=\frac{1}{64\pi}\left(1-\frac{m_\pi^2}{m_p^2}\right)^2\frac{m_p}{f^2} \ \frac{1}{2}(1+D+F)^2
\left\{
 \alpha_H^2A_{RL}^2 \frac{|\delta^L_{1111}|^2 }{M_{S_1}^4}
+\beta_H^2A_{RL}^2 \frac{|\delta^R_{1111}|^2 }{M_{S_1}^4}
 \right\},
\label{ptoepi}\\
\Gamma(p\to \bar{\nu}K^+)&=\frac{1}{64\pi}\left(1-\frac{m_K^2}{m_p^2}\right)^2\frac{m_p}{f^2}\ \alpha_H^2A_{RL}^2 \frac{1}{M_{S_1}^4}
\left| \delta^L_{2111} \frac{2D}{3} +\delta^L_{1122}\left(1+\frac{D}{3}+F\right)\right|^2,
\label{ptonuk}\\
\Gamma(p\to \bar{\nu}\pi^+)&=\frac{1}{64\pi}\left(1-\frac{m_\pi^2}{m_p^2}\right)^2\frac{m_p}{f^2} \ (1+D+F)^2\alpha_H^2A_{RL}^2 \frac{|\delta^L_{1111}|^2}{M_{S_1}^4},\label{ptomupi}
\end{align}
 where $m_p,m_K,m_\pi$ respectively denote the masses of proton, kaon, pion, $f$ denotes the pion decay constant, $D,F$ are parameters of the baryon chiral Lagrangian, $\alpha_H$ is the hadronic form factor, $A_{RL}$ accounts for renormalization group evolutions of the dimension-six operators.
For $m_p,m_K,m_\pi$, we use the values in Particle Data Group~\cite{pdg}.  We have $D=0.80$, $F=0.46$, $f=0.093$~GeV.
The hadronic form factor $\alpha_H$, which accounts for the non-perturbative strong interaction effects, is evaluated at the energy scale of 2 GeV using the result from Ref. [3], where $\alpha_H(2~{\rm GeV})=-\beta_H(2~{\rm GeV})=-0.0144$~GeV \cite{Aoki:2017puj}. This form factor plays a crucial role in capturing the non-perturbative dynamics of the proton decay process. Furthermore, the RGE of the dimension-six operators contributing to proton decay is taken into account through the factor $A_{RL}$. This factor is calculated by solving the one-loop RGE for the effective coupling of the proton decay operator, yielding a value of $A_{RL}=2.6$.

The mixing parameters $\delta^{L,R}_{ijkl}$, which encode the Yukawa couplings of the fermions, are defined as follows:
\begin{align}
\delta^L_{ijkl}=(Y_1^{DU})^*_{ij}(Y^{QL}_1)_{kl} ,
\nonumber \\
\delta^R_{ijkl}=(Y_1^{DU})_{ij}(Y^{UE}_1)_{kl}.
\end{align}
where $Y_1^{DU}$, $Y_1^{QL}$, $Y^{UE}_1$  are the Yukawa coupling matrices in the mass basis. These mixing parameters are evaluated at the energy scale $M_{S_1}$ by solving the corresponding RGEs. Notice that the proton decay mediated by the $Y_1^{QQ}$ coupling  is negligible since the $S_1$ scalar is predominantly composed of the 45-dimensional Higgs boson and  $Y_1^{QQ}$  is highly suppressed by the tiny mixing angle $\theta_S$.

To numerically calculate the proton decay rates, we follow a systematic procedure. First, we solve the RGEs for the SM Yukawa couplings, evolving them from the electroweak scale up to the GUT scale, $\Lambda = 6.5 \times 10^{15}~\mathrm{GeV}$. This GUT scale is determined by the reference values of the scalar masses given in Eq. \eqref{eq:spec}, which lead to successful GCU.
For the input parameters at the electroweak scale, we use the values listed in Table \ref{FermionMass}~\cite{pdg}. By solving the RGEs, we obtain the values of the Yukawa coupling matrices $Y_U$, $Y_D$, and $Y_E$ at the GUT scale.
From these Yukawa coupling matrices at the GUT scale, we can calculate the GUT-scale Yukawa couplings $Y^U_5$, $Y^D_5$, $Y^U_{45}$, and $Y^D_{45}$ using the matching equations in Eq. \eqref{eq:Ymatching2}. Importantly, we assume that the Yukawa coupling $Y^U_{45}$ vanishes at the GUT scale, following the assumption made in the previous discussion. This assumption implies that the unitary matrix $V_{QU}$ should be the identity matrix to satisfy the matching conditions. For the other unitary matrices, $V_{QE}$ and $V_{DL}$, we assume that they are random unitary matrices whose elements are distributed according to the Haar measure as in Ref.~\cite{Haba:2000be}, because their specific forms are not directly constrained by low-energy observables. We also assume that $\delta_H$ and $\theta_H$ are random values.
Once the GUT-scale Yukawa couplings are determined, we can obtain the Yukawa couplings for the light scalar components of the 45-Higgs, namely $S_8$, $S_3$, and $S_1$, from the matching conditions in Eqs. \eqref{eq:Ymatching}.
With these boundary conditions at the GUT scale, we solve the RGEs for the Yukawa couplings, including the contributions from the light scalar components of $45_H$, evolving them from the GUT scale down to the mass scale of $S_1$.
Using the Yukawa couplings at the $M_{S_1}$ scale, we can calculate the partial decay widths for various proton decay modes, such as $p \rightarrow \mu^+ K^0$ and $p \rightarrow \mu^+ \pi^0$, using the formulae given in Eqs. \eqref{ptomuk}-\eqref{ptomupi}. These partial decay widths depend on the Yukawa couplings, the scalar masses, and other input parameters related to the hadronic matrix elements and form factors.
If we further evolve these Yukawa couplings from the $M_{S_1}$ scale down to the electroweak scale using the RGEs, the resulting Yukawa coupling matrices $Y_U$, $Y_D$, and $Y_E$ would differ from their initial input values at the electroweak scale. This discrepancy arises because we first evolved the Yukawa couplings using the SM RGEs from the electroweak scale to the GUT scale, and then introduced additional contributions from the scalar components of $45_H$ when evolving from the GUT scale to the $M_{S_1}$ scale.
While this discrepancy exists, it is not expected to significantly affect the estimation of the proton decay rates, which primarily depend on the Yukawa couplings at the $M_{S_1}$ scale. For simplicity and to avoid unnecessary complications, we neglect this discrepancy and focus on the Yukawa couplings at the $M_{S_1}$ scale when calculating the proton decay rates.

\begin{table}[H]
\begin{center}
  \caption{The Yukawa coupling constants for the up-type quarks, down-type quarks, charged leptons and the Wolfenstein parameters of the CKM matrix at the $M_Z$ scale \cite{pdg}.}
  \label{values}
    \begin{tabular}{|c|c|c|c|c|c|c|c|c|} \hline
     $y_u$ & $y_c$ & $y_t$ & $y_d$ & $y_s$ & $y_b$&$y_e$&$y_\mu$&$y_\tau$\\\hline
     $7.05\times 10^{-6}$ &0.00363 &0.955 & $1.54\times 10^{-5}$  &0.000306 &0.0163
       & $2.79\times 10^{-6}$  &0.000590 &0.0100\\\hline\hline
      $\lambda$ & $A$ &$\overline{\rho}$ &$\overline{\eta}$ & & & & &  \\\hline
       0.225 & 0.826 & 0.159 &0.352  & & & & &  \\\hline     
  \end{tabular}
  \end{center}
           \label{FermionMass}
\end{table}

\begin{table}[h]
  \caption{Experimental constraints on the life times of various proton decay modes by Super-KamioKande \cite{Super-Kamiokande:2020wjk}\cite{Super-Kamiokande:2022egr}\cite{Super-Kamiokande:2014otb}\cite{Super-Kamiokande:2005lev}\cite{Super-Kamiokande:2013rwg}.}
  \begin{tabular}{|c|c|c|c|c|c|c|} 
   \hline 
   Decay mode & $p\rightarrow\mu^+K^0$ & $p\rightarrow\mu^+\pi^0$ & $p\rightarrow e^+K^0$ & $p\rightarrow e^+\pi^0$ & $p\rightarrow \overline{\nu}K^+$ & $p\rightarrow \overline{\nu}\pi^+$  \\\hline
	90\% C.L. (years) & $3.6\times10^{33}$ & $1.6\times10^{34}$ & $1.0\times10^{33}$ & $2.4\times10^{34}$ & $5.9\times10^{33}$ &$3.9\times10^{32}$ 
	\\\hline
  \end{tabular}
    \label{ProtonLifeTime}
\end{table}

\begin{figure}[H]
    \includegraphics[width=80mm]{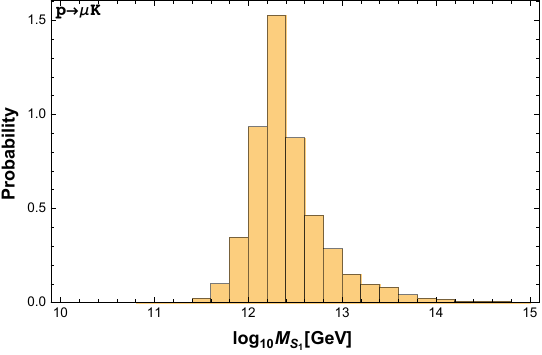}~~
    \includegraphics[width=80mm]{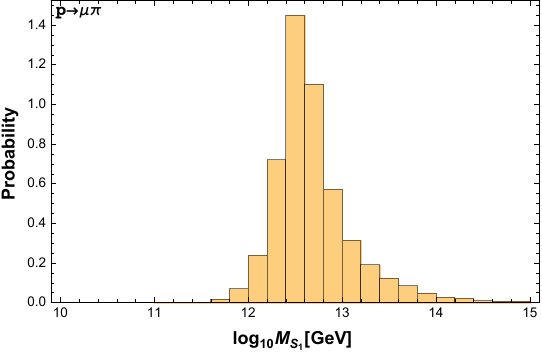}
\\

    \includegraphics[width=80mm]{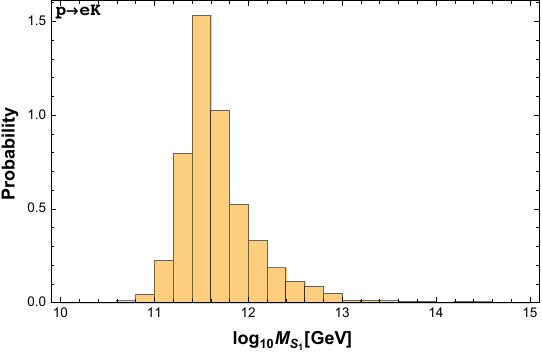}~~
    \includegraphics[width=80mm]{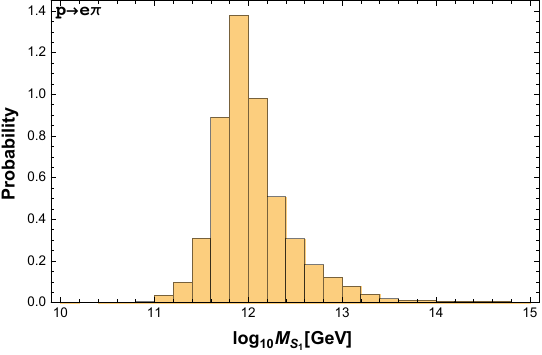}
\\
    \includegraphics[width=80mm]{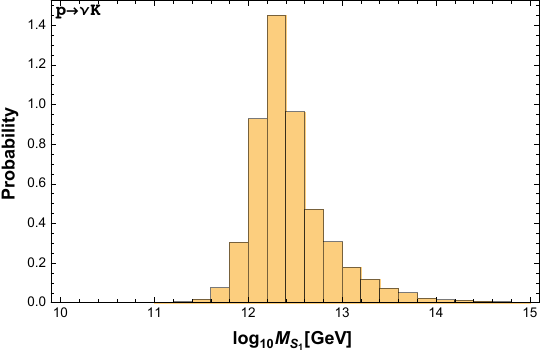}~~
    \includegraphics[width=80mm]{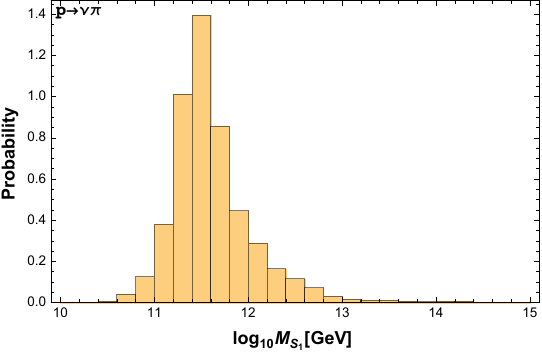}
  \caption{The probability distributions for the lower bound on the mass scale $M_{S_1}$ derived from the experimental constraints on various proton decay modes.
  }
  \label{protondecay1}
\end{figure}

In Fig.~\ref{protondecay1}, we present the probability distributions for the lower bounds on the mass $M_{S_1}$ derived from the experimental constraints on various proton decay modes,
which are listed in Table~\ref{ProtonLifeTime}.
Our analysis reveals that the $p \rightarrow \nu \pi$ mode yields the lowest peak position for the $M_{S_1}$ lower bound, around $3 \times 10^{11}$ GeV. On the other hand, The $p \rightarrow \mu \pi$ mode exhibits the highest peak position for the $M_{S_1}$ lower bound, approximately $3 \times 10^{12}$ GeV.  We found that the $p \rightarrow \nu \pi$ mode imposes the most stringent constraint on $M_{S_1}$.

\begin{figure}[H]
    \includegraphics[width=80mm]{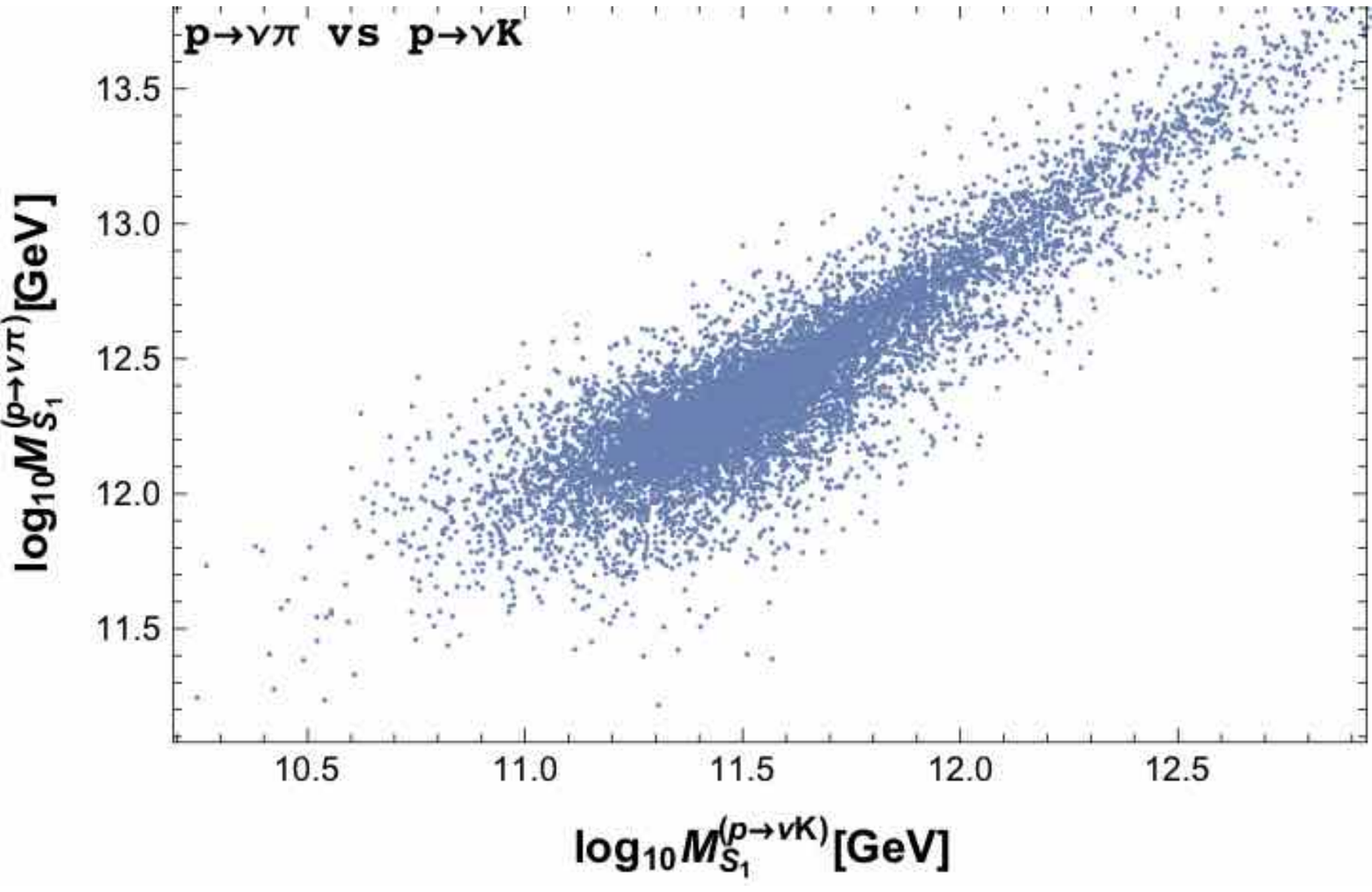}~~
    \includegraphics[width=80mm]{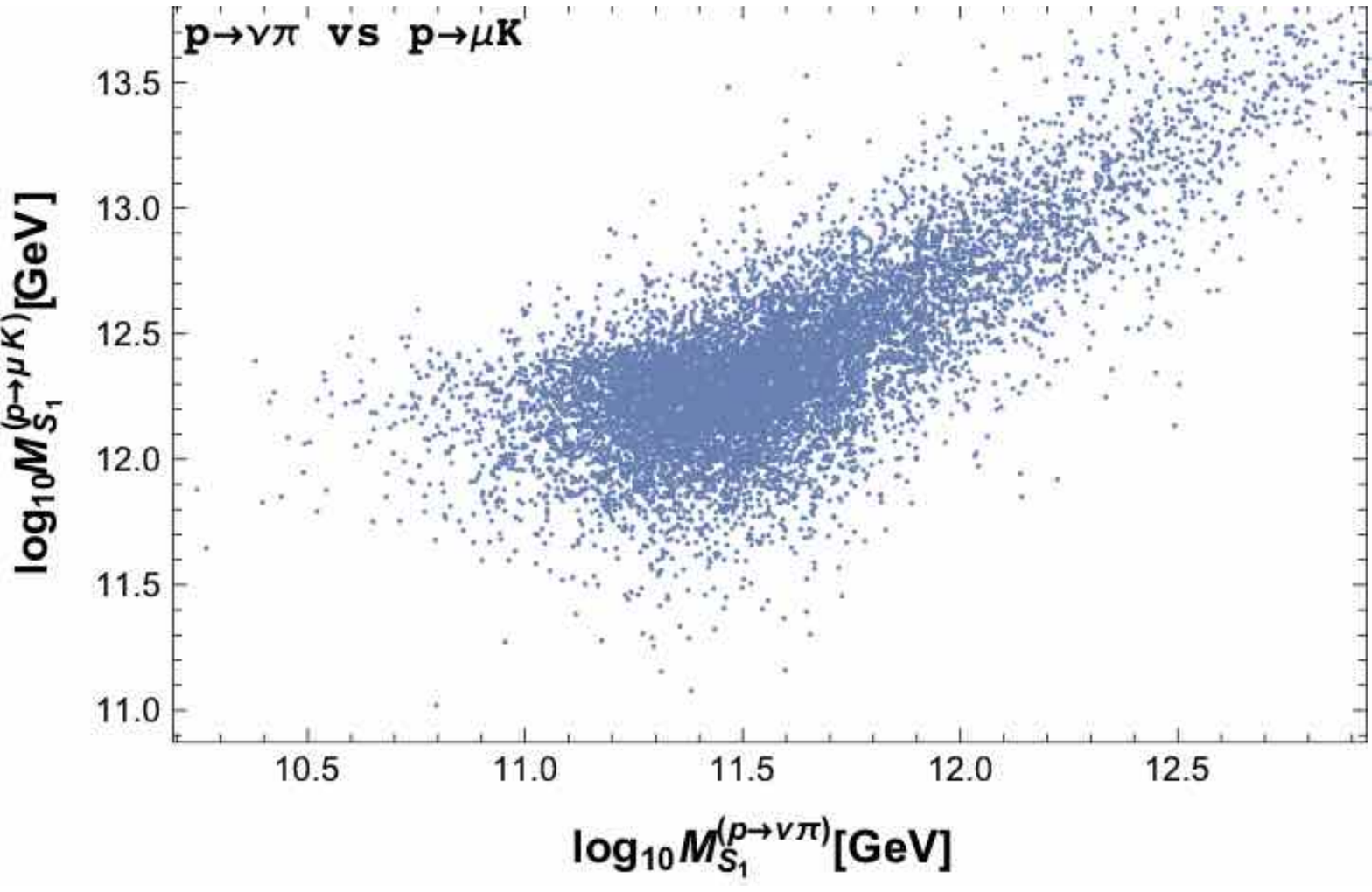}
\\

    \includegraphics[width=80mm]{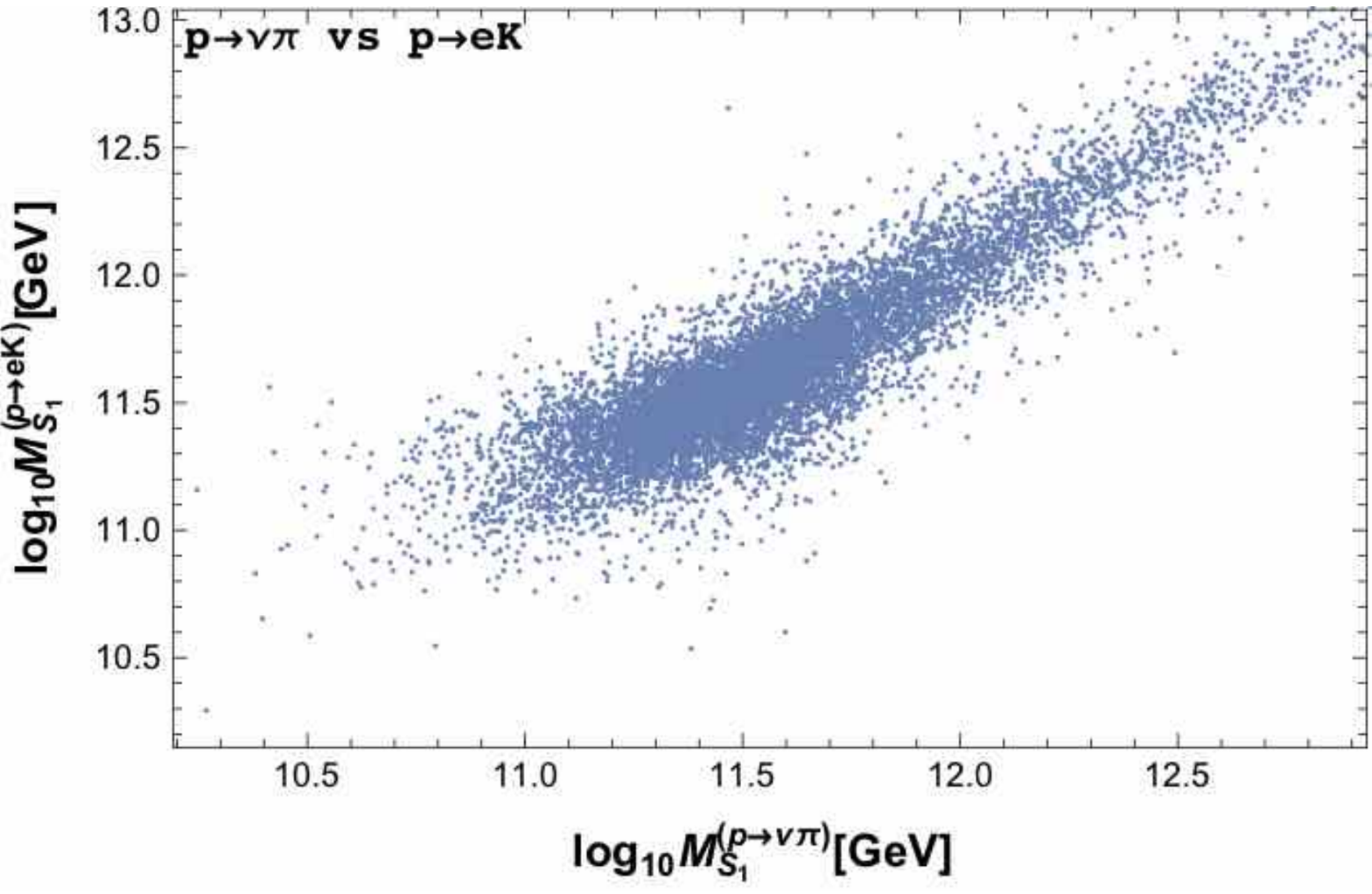}~~
    \includegraphics[width=80mm]{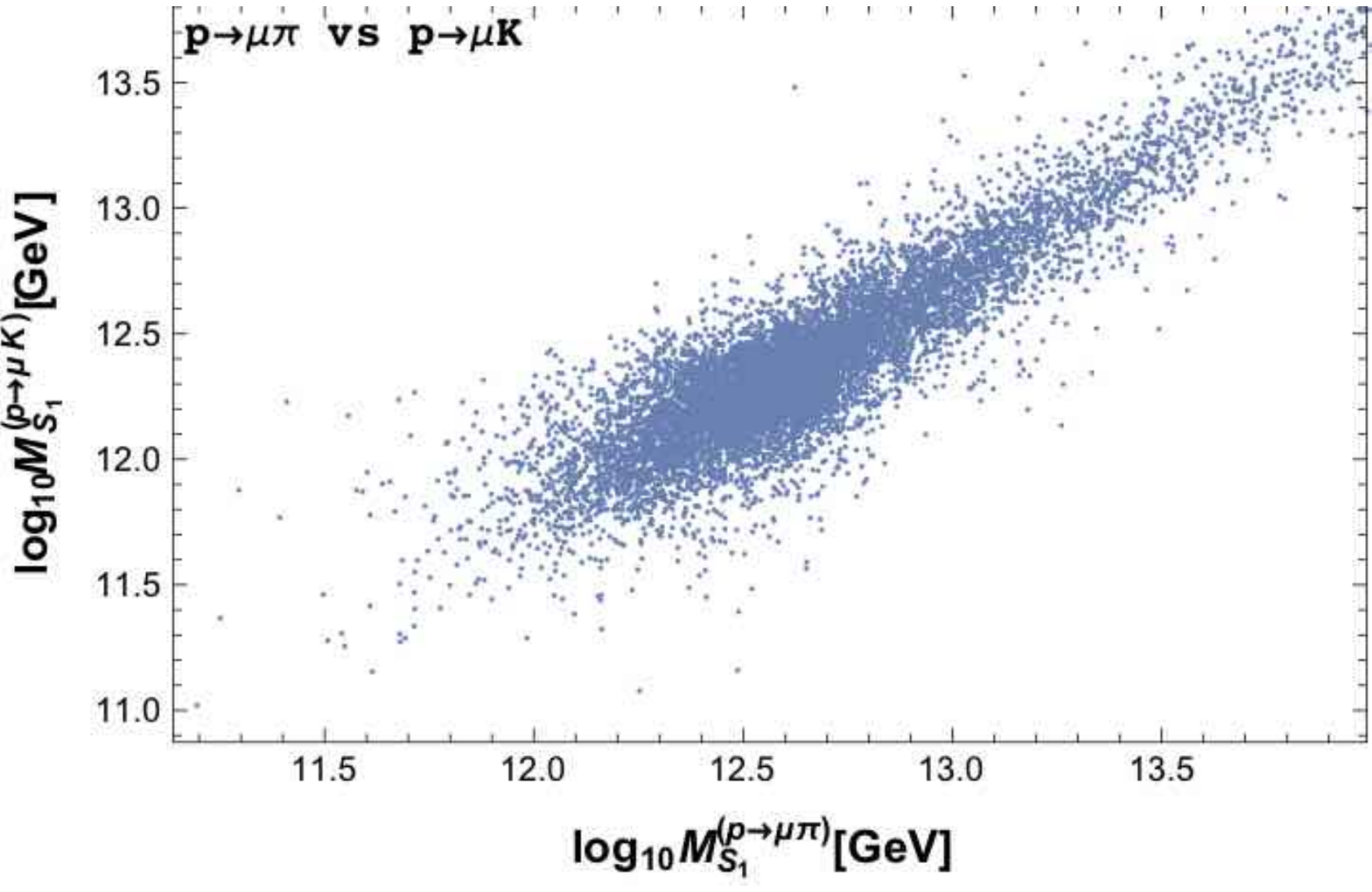}
\\
    \includegraphics[width=80mm]{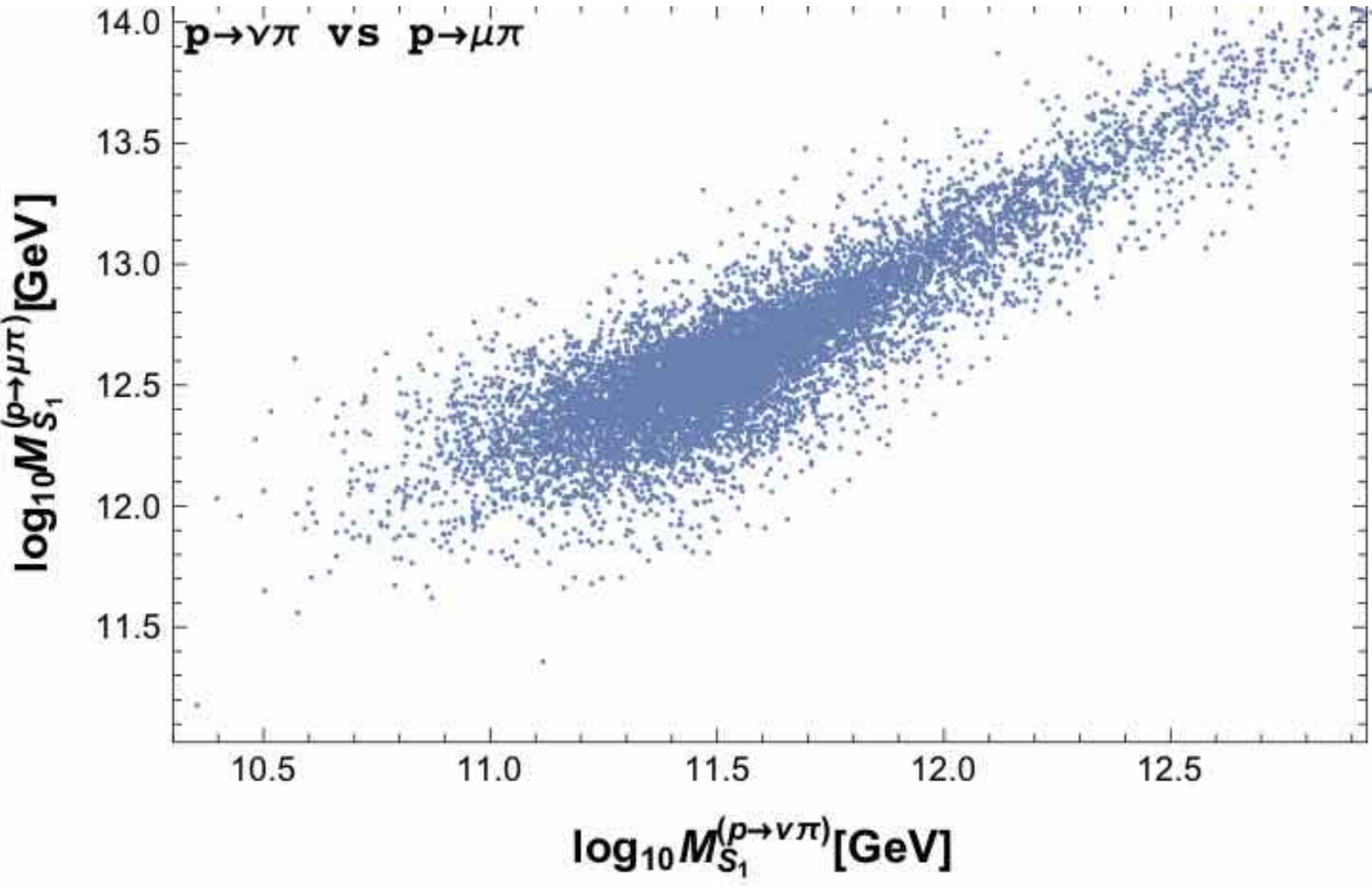}~~
    \includegraphics[width=80mm]{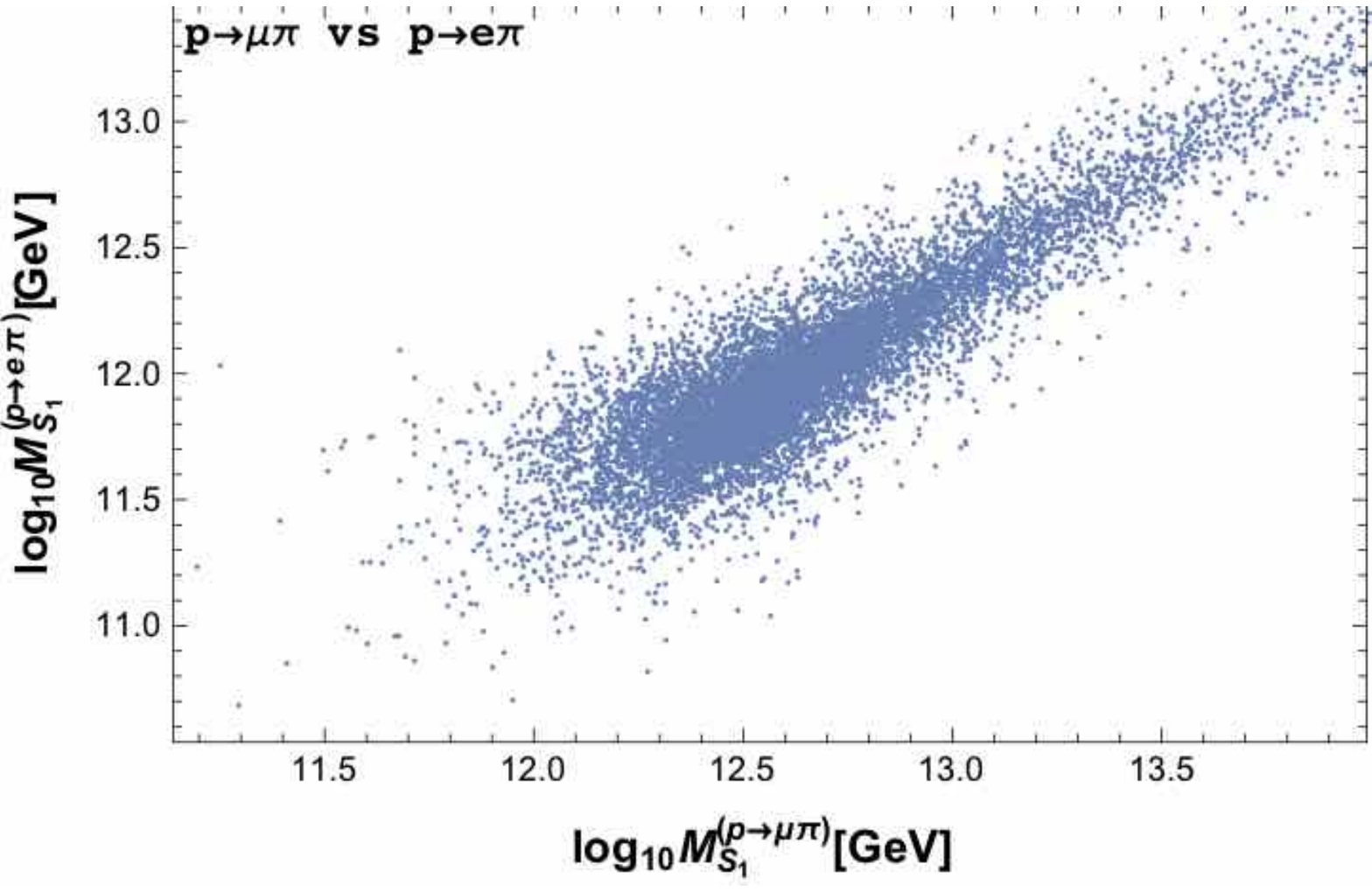}
  \caption{The correlations between the lower bounds on $M_{S_1}$ derived from various proton decay modes.
  }
  \label{protondecay2}
\end{figure}

In Fig.~\ref{protondecay2}, we present correlations among the lower bounds on the mass $M_{S_1}$ derived from the experimental constraints on various proton decay modes. 
We observe that the data points in the plot between $p \to \nu \pi$ and $p \to \mu K$ modes exhibit some scatter. This scatter can be attributed to the different contributions to the decay widths of these two modes.
The decay width for the $p \to \nu \pi$ mode is primarily determined by the term $|\delta^L|^2$, which represents the contribution from the left-handed mixing parameters $\delta^L_{ijkl}$. On the other hand, the decay width for the $p \to \mu K$ mode receives contributions from both $|\delta^L|^2$ and $|\delta^R|^2$ terms. The $|\delta^R|^2$ term represents the contribution from the right-handed mixing parameters $\delta^R_{ijkl}$. The presence of both $|\delta^L|^2$ and $|\delta^R|^2$ contributions in the $p \to \mu K$ decay width leads to a larger spread in the values of the lower bound on $M_{S_1}$ compared to the $p \to \nu \pi$ mode, which depends only on $|\delta^L|^2$. 
Furthermore, the left-handed mixing parameter for $p \to \mu K$, $\delta^L_{2112}=(Y_1^{DU})^*_{21}(Y^{QL}_1)_{12}$, involves the product of off-diagonal elements from two Yukawa matrices. This contrasts with other decay modes, where at least one element is diagonal. As off-diagonal elements tend to have a wider spread than diagonal elements, the $p \to \mu K$ mode exhibits the most significant scatter among all decay modes.
This difference in the dependence on the mixing parameters results in the observed scatter between the data points in the plot.
These plots demonstrate the importance of properly constraining these parameters using experimental data from different proton decay channels to obtain robust bounds on the mass scale $M_{S_1}$.

\section{Summary}
In this paper, we have studied proton decay in an extended SU(5) grand unified theory that incorporates a 45-dimensional Higgs representation. Our analysis demonstrates that the inclusion of $45$-Higgs provides a rich framework for addressing some of the shortcomings of the minimal SU(5) GUT, such as the inability to generate realistic fermion mass hierarchies and the lack of sufficient proton stability.

We have shown that successful gauge coupling unification can be achieved in this extended model by considering a hierarchical mass spectrum for the scalar components of $\Phi_{45}$. The color octet scalar ($S_8$), color triplet scalar ($S_3$), and color anti-triplet scalar ($S_1$) play a crucial role in realizing GCU when their masses are significantly lighter than the other components of $\Phi_{45}$. Furthermore, we have identified a wider region of the parameter space where GCU can be maintained, providing greater flexibility in accommodating experimental constraints.

Our proton decay analysis focused on the channels mediated by the exchange of the color anti-triplet scalar $S_1$. By calculating the proton decay rates using the Yukawa couplings obtained from renormalization group evolution and matching conditions at the GUT scale, we have explored the dependence of the decay rates on the model parameters. 
We found that the $p \to  \nu \pi$ mode imposes the most stringent constraint on $M_{S_1}$. 
The correlations between the lower bounds on the $S_1$ mass scale derived from different proton decay modes, particularly the $p \to \nu \pi$ and $p \to \mu K$ channels, reveal the intricacies of the model's parameter space.

The scatter observed in the correlation plot between the $p \to \nu \pi$ and $p \to \mu K$ modes highlights the sensitivity of the proton decay rates to the Yukawa couplings and mixing parameters of the model. This underscores the importance of precisely measuring these parameters and using experimental data from various proton decay channels to place robust constraints on $M_{S_1}$.

\section*{Acknowledgments}
This work is partially supported by Scientific Grants by the Ministry of Education, Culture, Sports, Science and Technology of Japan, No. 21H00076 (NH) and No. 19K147101 (TY).

%%%%%%%%%%%%%%%%%
%%% References %%%
%%%%%%%%%%%%%%%%%

\end{document}